\documentclass[3p,times,twocolumn]{elsarticlenohead}
\usepackage{amssymb}
\usepackage[figuresright]{rotating}
\usepackage{amsmath}
\usepackage{fixltx2e}

\providecommand{\proarrow}[0]{\rightarrow}

\providecommand{\dif}[0]{\mathrm{d}}

\providecommand{\proname}[2]{#1 \proarrow #2}
\providecommand{\lrproname}[2]{#1 \leftrightarrow #2}

\providecommand{\abss}[1]{\left\lvert #1 \right\rvert^2}

\providecommand{\order}[1]{O\left( #1 \right)}
\providecommand{\torder}[1]{O\bigl( #1 \bigr)}

\providecommand{\ldl}[1]{(\lambda^\dag \lambda)_{#1 #1}}

\providecommand{\ghor}[2]{\gamma \left(\proname{#1}{#2}\right)}

\providecommand{\mplanck}[0]{m_{\rm P}}
\begin{document}

\begin{frontmatter}

\title{Mini-review on baryogenesis at the TeV scale and possible connections with dark matter}

\author{J. Racker}

\address{Departamento\ de F\'{\i}sica Te\'orica, Instituto de F\'isica corpuscular (IFIC), Universidad de
Valencia-CSIC \\ 
Edificio de Institutos de Paterna, Apt. 22085, 46071 Valencia,
Spain}

\begin{abstract}
This is a very short review on different mechanisms for baryogenesis from particle decays or annihilations at low temperature ($T \lesssim 10$~TeV) 
and their implementation in models that relate the origin of the baryon asymmetry and dark matter.
\end{abstract}

\begin{keyword}
Baryogenesis \sep Dark matter \sep  Neutrino masses \sep Beyond Standard Model
\end{keyword}

\end{frontmatter}

Many scenarios for explaining the origin of the Baryon Asymmetry of the Universe (BAU), notably ``standard'' leptogenesis, must take place at high temperatures ($T \gtrsim 10^8$~GeV). However there are several motivations to explore models for baryogenesis at lower scales, like accessibility to experimental exploration, compatibility with some well motivated supergravity models that require reheating temperatures below $10^5-10^7$~GeV not to overproduce gravitinos~\cite{Khlopov84,Ellis84}, and avoidance of hierarchy problems~\cite{vissani97}. Moreover, it could also happen that baryogenesis above a few TeVs becomes severely disfavored, demanding a low-scale explanation for the BAU, e.g. if some lepton number ($L$) violating processes are observed at the LHC~\cite{frere08,deppisch13}. Here  we review the main problems to achieve baryogenesis from heavy particle decays or annihilations at low temperatures, mechanisms to reach the TeV scale, and some examples of their use in models connecting the origin of the BAU and Dark Matter (DM).  

\begin{figure*}[!t]
\begin{center}
\includegraphics[width=1.9\columnwidth]{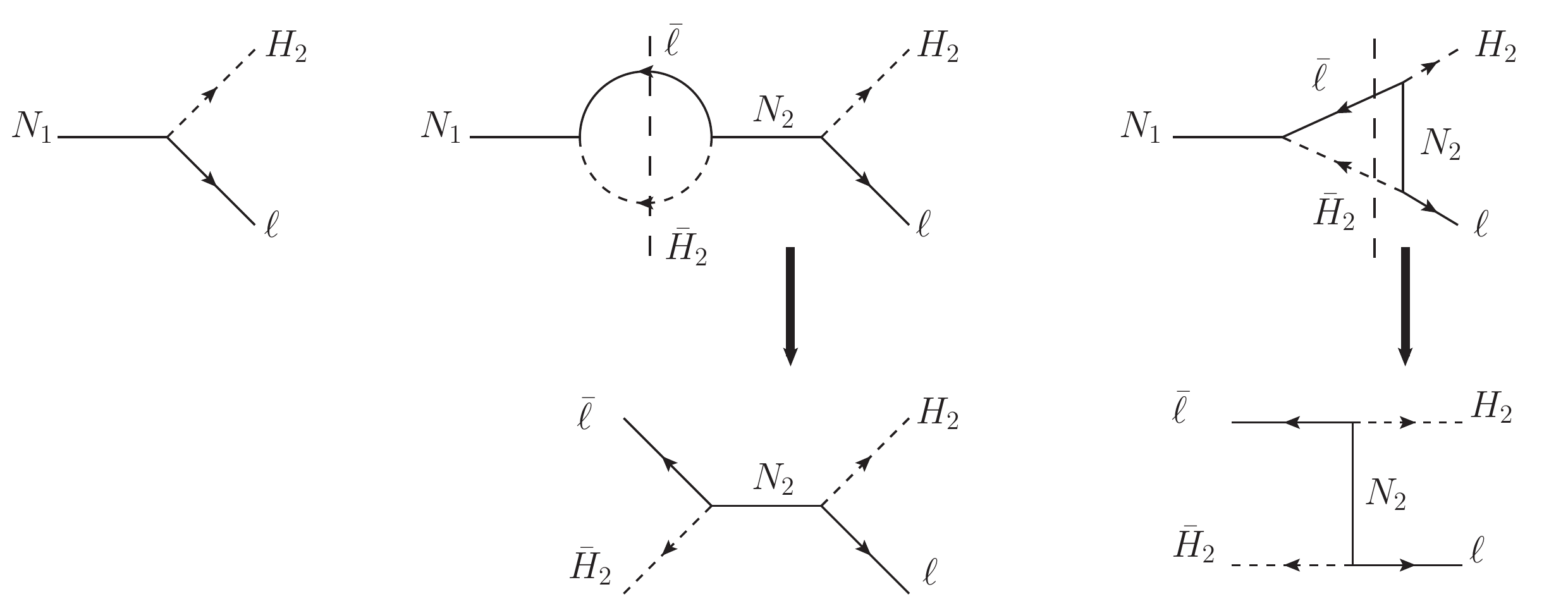}
\end{center}
\caption{The CP asymmetry in $N_1$ decays comes -at leading order- from the interference of the tree-level and one-loop diagrams shown on the top. The vertical cuts through the loops indicate that those particles can go on shell, which is necessary to have a CP even phase. In turn, this implies the existence of the processes on the bottom, which contribute to the washout of the lepton asymmetry.}
\label{fig:cpasym}
\end{figure*}

A crucial quantity for baryogenesis is the amount of CP violation per each decay or annihilation, to be denoted by $\epsilon$. E.g. in type I unflavoured leptogenesis, $\epsilon \equiv \tfrac{\ghor{N}{\ell H} - \ghor{N}{\bar \ell \bar H}}{\ghor{N}{\ell H} + \ghor{N}{\bar \ell \bar H}}$, where $N$ is a sterile neutrino, $H$ the Standard Model (SM) Higgs, $\ell$ a SM lepton doublet, and the $\gamma$'s denote number of processes per unit time and volume. To simplify this discussion we take $\epsilon$ independent of the temperature (which is often a very good approximation). In this case the baryon density asymmetry, $n_B$, conveniently normalized to the entropy density, $s$, is directly proportional to $\epsilon$ and can be expressed as $Y_B \equiv \tfrac{n_B}{s} = \kappa \epsilon \eta$. The ``efficiency'' $\eta$ depends on the kinetics of baryogenesis, while the numerical coefficient $\kappa$ is defined in each model under the condition that the maximum efficiency be 1. There are two main problems to have baryogenesis at low scales: 
\begin{enumerate}[(a)]
\item In some leptogenesis models the CP asymmetry $\epsilon$ is proportional to the masses of both, the light -mostly active- neutrinos and the heavy ones~\cite{davidson02}. Given the tiny value of the light neutrino masses, the mass $M$ of the heavy neutrinos -which sets the scale for baryogenesis- must be very high to have a large enough $\epsilon$. E.g. in type I leptogenesis with hierarchical heavy neutrinos, $M \gtrsim 10^8 - 10^9$~GeV~\cite{davidson02,hambye03}. This fact seems to be well known. 
\item The second problem is intrinsic to all scenarios of baryogenesis from particle decays or annihilations, but curiously enough it is often either completely ignored or not treated properly. CP violation requires a CP odd and a CP even phase. The former arises from complex couplings, the later from the absorptive part of one loop contributions. In turn the absorptive part is proportional to the amplitude of processes which violate baryon number ($B$) -or $L$-~\cite{nanopoulos79}, and whose strength, therefore, cannot be very small 
for the CP asymmetry to be large enough.
If baryogenesis occurs at low temperatures, when the expansion rate of the universe is mild, these processes are typically very fast and washout the asymmetry.
\end{enumerate}
To illustrate problem (b) consider leptogenesis in a radiative seesaw model -to be safe from problem (a)-~, namely the Inert Doublet Model (IDM) with singlet neutrinos: The SM is extended with a Higgs doublet, $H_2$, and at least two heavy sterile neutrinos, $N_i \, \{ i=1,2\}$. These extra fields are the only ones odd under a new $Z_2$ discrete symmetry, which is assumed to be exactly conserved, so that $H_2$ does not acquire a vacuum expectation value~\cite{ma06}. Choosing the mass of $N_1$ larger than the mass of $H_2$, $M_1 > M_{H_2}$, the CP asymmetry in $N_1$ decays arises from the interference of tree level an one loop diagrams, as depicted in figure~\ref{fig:cpasym}. The CP asymmetry $\epsilon$ is proportional to the amplitude of the processes at the right of the cut, $\lrproname{H_2 \ell}{\bar H_2 \bar \ell}$~\footnote{To simplify the notation we will often omit flavour indices.}, which may washout the lepton asymmetry depending on how fast they are compared to the Hubble rate $H$. Since $H \propto T^2/m_P$, with $T$ the temperature, there is an absolute energy scale given by the Planck mass $\mplanck$: the lower the scale of baryogenesis compared to $\mplanck$, the harder it becomes to have at the same time a large enough CP asymmetry and small washouts. 

\begin{figure}[!htb]
\begin{center}
\includegraphics[width=0.65\columnwidth,angle=270]{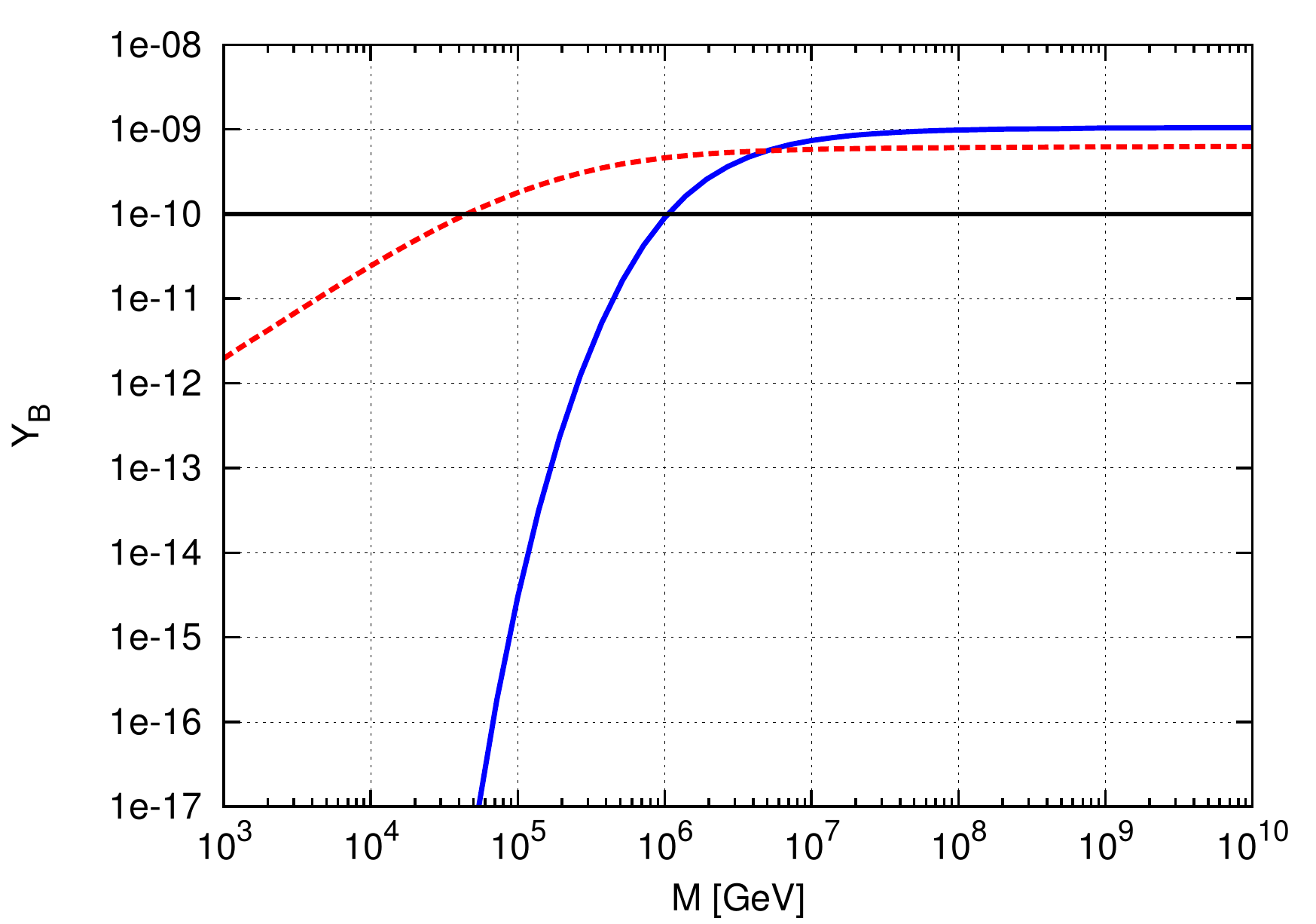}
\end{center}
\caption{The final baryon asymmetry, $Y_B$, as a function of $M$, including the washout processes $\lrproname{H_2 \ell}{\bar H_2 \bar \ell}$ and $\lrproname{\ell \ell}{\bar H_2 \bar H_2}$ (solid blue curve) or not including them (dashed red curve). The Yukawa couplings and $M_2$ are chosen so that $\frac{\Gamma_1}{H(T=M)}$ and $\epsilon$ keep constant for all $M$, where $\Gamma_1$ is the decay width of $N_1$.
As can be seen, for $M \lesssim 10^7$~GeV the baryon asymmetry falls exponentially with decreasing $M$. Instead, if the washouts are -incorrectly- not included, $Y_B$ only declines linearly and its value for $M$ at the TeV scale is wrong by many orders of magnitude.}
\label{fig:washout}
\end{figure}

In the IDM it was shown~\cite{racker13} that this intimate connection between CP asymmetry and washouts precludes leptogenesis for, roughly, $M \equiv M_1 \lesssim \torder{10^2}$~TeV, unless one of the low-scale-baryogenesis mechanisms described below is used. Arguably this bound is quite model independent~\cite{racker13} (within an order of magnitude, check e.g.~\cite{racker12} for the analogous result in the inverse seesaw~\footnote{A quite different scenario involving CP-violating decays and annihilations has been presented in~\cite{baldes14}. Although the parameter space was not completely scanned, the results also suggest a bound $M \gtrsim \torder{10^2}$~TeV for the scale of baryogenesis (when none of the low-scale-baryogenesis mechanisms described below are implemented).}). The importance of this type of washouts also becomes apparent comparing the resultant baryon asymmetry after including them or not, see figure~\ref{fig:washout}. 

Next we briefly describe the three mechanisms which, in our opinion, have been demonstrated to allow for baryogenesis at the TeV scale -or even below-.
\begin{itemize}
\item {\it Almost degenerate particles}: When there is a pair of almost degenerate particles, the CP asymmetry in decays can be enhanced up to $\order{1}$ values~\cite{flanz96,covi96II}. To be more concrete consider two sterile neutrinos $N_{i}\, \{i=1,2\}$, with masses $M_{i}$, Yukawa couplings $\lambda_{\alpha i}$ with the SM lepton doublet $\ell_\alpha$, and decay widths $\Gamma_i$. If the resonant condition $\Delta M \equiv M_2 - M_1 = \Gamma_2/2$ holds, then the CP asymmetry in $N_1$ decays, $\epsilon_1$, can take values as large as 1/2 independently of the size of the Yukawa couplings of $N_2$, i.e. of  $\ldl{2}$. Moreover, for $\Gamma_{1,2} \ll \Delta M \ll M_1$, $\epsilon_1 \propto \ldl{2}/\delta$, with $\delta \equiv \Delta M/M_1$.  
Therefore it is possible to reduce the ``dangerous'' washouts taking $\ldl{2}$ small enough, while keeping $\epsilon_1$ sizable by choosing a tiny value for $\delta$. This -so called resonant leptogenesis- mechanism has been widely studied (see e.g.~\cite{pilaftsis03}) and it can lead to successful baryogenesis at the TeV scale. The amount of degeneracy required has been studied in detail in~\cite{racker13} for the case of sterile neutrinos. It depends on $M_1$ and on the hierarchy among the Yukawa couplings, $r$, defined as the minimum non-null quantity in the list $\left\{\sqrt{\ldl{1}/\ldl{2}}, \sqrt{K_{\alpha i}} \; (i=1, 2;\; \alpha=e, \mu, \tau)\right\}$, where the projectors are given by $K_{\alpha i} \equiv \tfrac{\abss{\lambda_{\alpha i}}}{(\lambda^\dag \lambda)_{ii}}$. E.g. the following rules of thumb apply to get the BAU via the resonant leptogenesis mechanism: 
\begin{equation*}
\label{eq:rule}
\begin{split}
&\delta \cdot r \lesssim 10^{-8}\; {\rm for}  \; M_1 \sim 4~{\rm TeV}\;, \\
&\delta \cdot r \lesssim 2 \times 10^{-9}\; {\rm for} \; 250~{\rm GeV} \lesssim M_1 \lesssim 1~{\rm TeV}\;.
\end{split}
\end{equation*}
\item  {\it Late decay}: The rate of the washout processes associated to the CP asymmetry, like $\lrproname{H_2 \ell}{\bar H_2 \bar \ell}$ in figure~\ref{fig:cpasym}, is proportional to $(T/M_2)^a$ when $T \lesssim M_2$, with $M_2$ the mass of the mediator and $a=2\, (4)$ for a Majorana (Dirac or scalar) propagator. Increasing the value of $M_2$ relative to $M_1$ (the mass of the decaying particle $N_1$) does not help, of course, to lower the scale of baryogenesis, because the CP asymmetry and washouts would be reduced by the same amount. However the later the $N_1$'s decay, the smaller the washouts will be at the time the baryon (or lepton) asymmetry is generated. Successful baryogenesis at the TeV scale via this late-decay scenario requires two conditions: a small decay width $\Gamma_1 \ll H(T=M_1)$ and a new process to create the $N_1$'s (given that the inverse decays and related scatterings are tiny), e.g. an exotic gauge interaction~\cite{plumacher96,racker08}. Note that the ``new'' process must decouple before $N_1$ decays, so that the $N_1$'s disappear via the CP-violating interaction. In this way it is possible to have leptogenesis at $T \sim$ few~TeVs (the exact value depending on the freeze-out temperature of the sphalerons) and baryogenesis at much lower scales if $B$ is violated perturbatively.
\item {\it Massive decay or annihilation products}: It has been explained that the CP asymmetry is proportional to the amplitude of processes that violate $B$ (or $L$) and which therefore contribute to the washout of the baryon asymmetry. The rate of the washouts is not only determined by the amplitude of the corresponding process, but also by the number density of the particles involved in the interaction. If at least one of these particles is massive and becomes non-relativistic during the baryogenesis epoch, the corresponding washout rate will be Boltzmann suppressed. This mechanism was shown to allow for TeV scale baryogenesis from DM annihilation first in~\cite{cui11} and from heavy particle decays in~\cite{racker13}. 

Actually the full mechanism is somewhat more complicated. To see why let us consider again the IDM as an example, with $H_2$ playing the role of the massive decay product. The decays of $N_1$ generate an asymmetry not only in leptons, but also in $H_2$. It is a simple statistical fact that the asymmetries in both sectors, $\ell$ and $H_2$, induce washouts of the lepton asymmetry (e.g. more $\ell$'s than $\bar \ell$'s, or more $H_2$'s than $\bar H_2$'s, favors the occurrence of $\proname{\ell H_2}{\bar \ell \bar H_2}$ over $\proname{\bar \ell \bar H_2}{\ell H_2}$). At linear order in the asymmetries $Y_{\Delta X} \equiv Y_X - Y_{\bar X} \equiv \tfrac{n_X}{s} - \tfrac{n_{\bar X}}{s}$ (and assuming kinetic equilibrium),
\begin{equation*}
\label{eq:wo}
\frac{\dif Y_{\Delta_\ell}}{\dif z} = \frac{-1}{sHz} \left\{ \frac{Y_{\Delta_\ell}}{Y_\ell^{\rm eq}} + \frac{Y_{\Delta H_2}}{Y_{H_2}^{\rm eq}} \right\} \gamma^{\rm eq}\left(\lrproname{\ell H_2}{\bar \ell \bar H_2}\right)+ \dots \; ,
\end{equation*}
where $z \equiv M_1/T$ and the superscript ``eq'' indicates use of equilibrium phase space distributions with zero chemical potentials. To appreciate the washout effects on $Y_{\Delta_\ell}$, write the ith term of the RHS of this equation as $-Y_{\Delta \ell} \, w_i(z)$ and analyze the function $w_i(z)$. The first term corresponds to a washout $\propto \gamma^{\rm eq}(\lrproname{\ell H_2}{\bar \ell \bar H_2})/Y_{\ell}^{\rm eq}$, which is clearly Boltzmann suppressed by $e^{-M_{H_2}/T}$. The problem arises with the second one: how is $Y_{\Delta H_2}$ related to $Y_{\Delta \ell}$? If they were linked by a conservation law, so that $Y_{\Delta H_2} = {\rm number} \times Y_{\Delta \ell}$, then $w_2(z) \propto \gamma^{\rm eq}(\lrproname{\ell H_2}{\bar \ell \bar H_2})/Y_{H_2}^{\rm eq}$, which is NOT Boltzmann suppressed (as long as the mass of $\ell$ is negligible). In this case the mechanism does NOT work to lower the scale of baryogenesis. Instead 
if $H_2$ can disappear through some fast interaction different from the one at the origin of the BAU, the corresponding chemical equilibrium condition would imply that $Y_{\Delta H_2}$ is Boltzmann suppressed with respect to $Y_{\Delta \ell}$. Then it is possible to achieve leptogenesis at the TeV scale (or, if $B$ is violated perturbatively, baryogenesis at much lower temperatures). 

The IDM is very useful to illustrate this issue. One of the allowed terms in the scalar potential is $\lambda_5 (H^\dag H_2)^2/2$. When the coupling $\lambda_5 = 0$ a conserved ``lepton'' number can be defined, assigning $L=-1$ to $H_2$. Instead for large $\lambda_5$ there are fast ``lepton'' number violating interactions, such as $\lrproname{H_2 \bar H}{\bar H_2 H}$. The amount of baryon asymmetry that can be generated at the TeV scale as a function of $\lambda_5$ is plotted in figure~\ref{fig:lam5}, which should serve as a lesson of when it is possible and when it is not possible to have baryogenesis at the TeV scale using massive decay or annihilation products. Later we will review some variations on this mechanism exemplified with models of baryogenesis from DM annihilation.
\end{itemize} 
\begin{figure}[!thb]
\begin{center}
\includegraphics[width=0.65\columnwidth,angle=270]{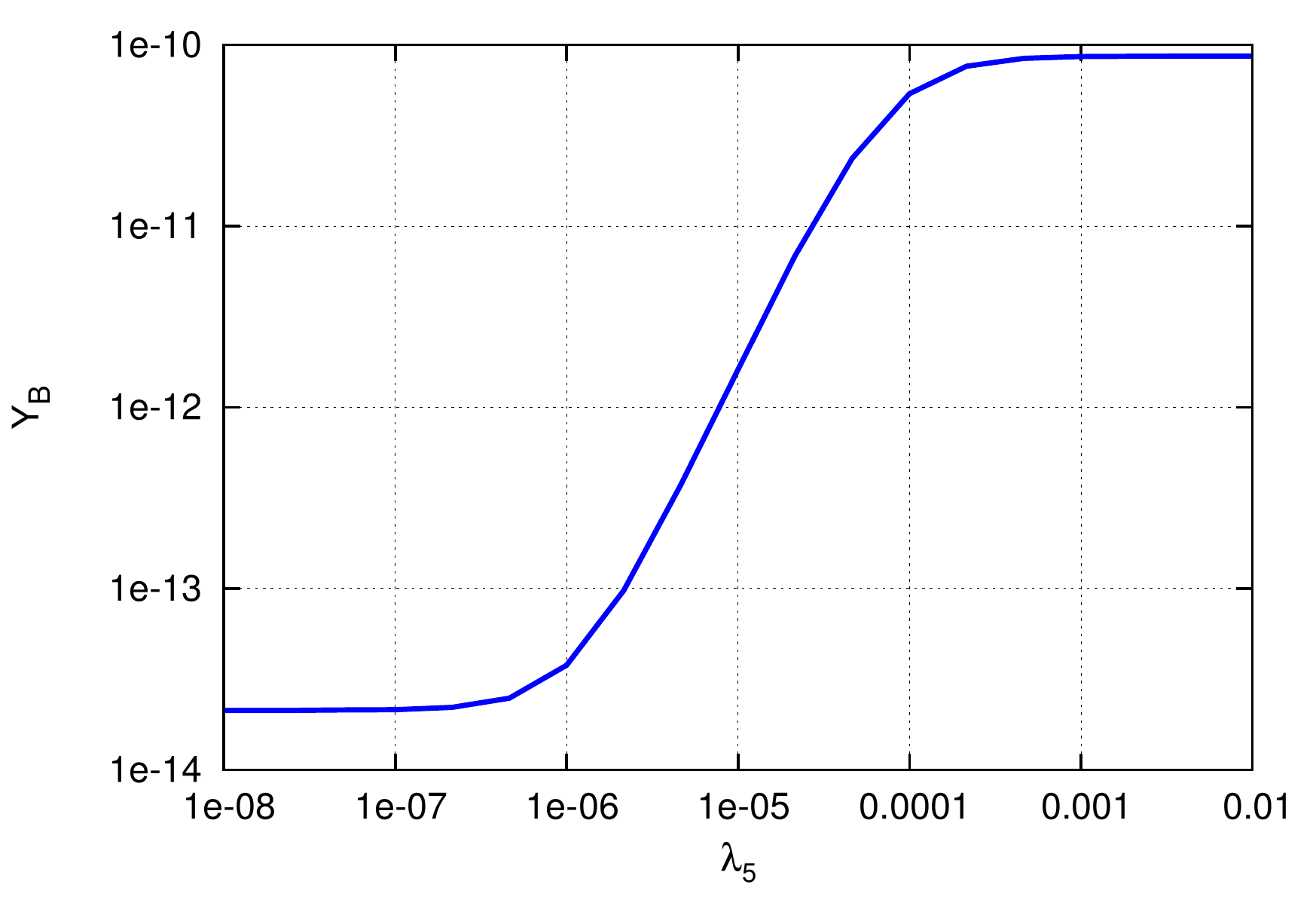}
\end{center}
\caption{The final baryon asymmetry, $Y_B$, as a function of $\lambda_5$ in the IDM with $M_1=2.8$~TeV and $M_{H_2}=0.6 M_1$. The freeze-out temperature of the sphalerons is taken equal to $T_* \simeq 130$~GeV~\cite{donofrio14}. The Yukawa couplings and $M_2$ have a fix value which is not important for the discussion. It is clear that the BAU, $Y_B^{\rm obs} \sim 10^{-10}$, can only be obtained for large values of $\lambda_5$, so that the processes $\lrproname{H_2 \bar H}{\bar H_2 H}$ are fast during the relevant period for baryogenesis.}
\label{fig:lam5}
\end{figure}

We have seen that the IDM is a simple extension of the SM which allows to incorporate any of the three low-scale-baryogenesis mechanisms described above~\footnote{The realization of the late-decay scenario requires some extra way of producing the heavy neutrinos.}~\cite{racker13} and it also provides a DM candidate, $H_2$. A similar situation occurs when extending the SM with three sterile neutrinos, $N_i$, and a charged scalar, $\delta^+$, singlet under SU(2)~\cite{frigerio14} (in this case the DM is $N_1$, with $M_1 >$ few~KeVs and the BAU can be generated from the CP-violating decays of $N_2$). To finish this very short review we wish to quote an example, for each of the three mechanisms, of their implementation in models aimed to explain the similar abundances of baryonic and dark matter (at least up to some degree) and that preserve the ``WIMP miracle''.

In~\cite{davidson12} baryogenesis occurs via leptogenesis in the decay of some gauge singlet Dirac fermions $\Psi$ with $\order{1}$~TeV masses. The density of $\Psi$ when the baryon asymmetry starts to build up has been established by the freeze out of annihilations, so that the baryon density asymmetry is similar to the relic density of a Weakly Interacting Massive Particle (WIMP) if the CP asymmetry is $\order{1}$. Then a pair of almost degenerate $\Psi$'s are required to have a large CP asymmetry and leptogenesis at the TeV scale.

The late decay mechanism was incorporated in~\cite{cui12}, where the scale of baryogenesis is only constrained to be above that of Big Bang nucleosynthesis. In this scenario there are at least two species of WIMPs, $\chi_{\text{DM}}$ which is stable DM, and $\chi_B$ which is meta-stable and decays after its annihilations freeze out. The decay violates CP and $B$, generating an amount $\epsilon$ of baryon asymmetry per decay. Since the $\chi_B$'s start to decay at $T \ll m_{\chi_B}$, baryogenesis is safe from the type of dangerous washouts depicted in figure~\ref{fig:cpasym}. Moreover, the population of $\chi_B$'s has a WIMP-like relic density at the start of baryogenesis, which was established by the freeze out of interactions different from the CP-violating ones. Altogether, there is a simple relation among the relic abundances of baryons ($\Omega_B$) and DM ($\Omega_{\rm DM} = \Omega_{\chi_{\rm DM}}$):
\begin{equation*}
\frac{\Omega_B}{\Omega_{\rm DM}} \simeq \epsilon \frac{m_p}{m_{\chi_B}} \frac{\tilde\Omega_{\chi_B}}{\Omega_{\chi_{\rm DM}}} \; ,
\end{equation*}
where $m_p$ is the proton mass, $m_{\chi_B}$ the $\chi_B$ mass, and $\tilde\Omega_{\chi_B}$ the would-be relic abundance of $\chi_B$ were it stable.

Finally let us consider baryogenesis from DM annihilations, which is a quite minimalistic way of relating the BAU and DM while preserving the WIMP miracle, because the baryon asymmetry is generated in the same annihilations whose freeze out determine the abundance of DM. Here a massive annihilation product $\Psi$ is used to allow for baryogenesis at a low scale -set by the mass of the DM, $m_\chi$-. Then the DM $\chi$ annihilates via $\proname{\chi \chi}{\Psi a,\, \bar\Psi \bar a}$, where $a$ is some light -or massless- field which accumulates particle-antiparticle asymmetry due to the CP-violating annihilations. This asymmetry might be baryonic or of some other type (like leptonic), in which case it must be transferred to the baryon sector by some fast process (like the electroweak sphalerons). In turn the asymmetry in the heavy $\Psi$'s could induce fatal washouts, as explained before. To avoid this problem different ideas have been proposed. 

In~\cite{cui11} $\Psi$ is an exotic vector-like lepton or quark that decays due to fast interactions involving light sterile particles (see also~\cite{bernal12}). These particles belong to a dark sector decoupled from the SM fields at low temperatures. Both models with and without a conserved $B-L$ were presented in~\cite{cui11} (in the former case the asymmetry in $\Psi$ is transferred to the light sterile sector avoiding a complete cancellation of the baryon asymmetry after $\Psi$ decays). Then in~\cite{bernal13} it was shown that successful baryogenesis is possible without the light hidden sector, even in models with a conserved $B-L$. Here $\Psi$, a vector-like quark, decays fast into an exotic particle $\Psi_2$, that in turn decays into SM particles after the electroweak sphalerons freeze out (hence the final baryon asymmetry is equal to the lepton asymmetry that keeps frozen after the sphalerons decouple). 

A different approach to solve the problem associated with the asymmetry in the heavy annihilation (or decay) product was proposed in~\cite{racker14}, namely to make a Majorana fermion or real scalar play the role of $\Psi$. In this way there is no asymmetry accumulating in the heavy field and the aforementioned problem is completely avoided. An implementation of this idea developed in~\cite{racker14} is to have baryogenesis from DM annihilation, like in the models just described, but the DM annihilates via $\proname{\chi \chi}{N \nu}$, where $N$ is a heavy Majorana neutrino with $M = \order{1}$~TeV. The DM $\chi$ is a singlet fermion and therefore $\nu$ must also be a sterile neutrino (with mass $m_\nu$). As long as $m_\nu \ll T$, an helicity asymmetry might be generated in the annihilations. The helicity asymmetry can be partially transferred to the SM lepton sector through fast Yukawa interactions and subsequently to baryons via the electroweak sphalerons. Therefore the BAU freezes in when the sphalerons decouple at $T_* \sim 100$~GeV and the condition for $m_\nu$ becomes $m_\nu \ll T_*$. In this model the sterile neutrinos are responsible for the masses of the SM neutrinos and they play a key role in baryogenesis and the freeze out of DM annihilations.     
\section*{Acknowledgments}
I wish to thank my collaborators on this subject, Nicol\'as Bernal, Stefano Colucci, Fran\c{c}ois-Xavier Josse-Michaux, Nuria Rius, and Lorenzo Ubaldi.

This work has been supported by the Spanish MINECO Subprogramme Juan de la Cierva and it has also been partially supported by the Spanish MINECO grants FPA2011-29678-C02-01 and Consolider-Ingenio CUP (CSD2008-00037), and by Generalitat Valenciana grant PROMETEO/2009/116. In addition we acknowledge partial support from the  European Union FP7  ITN INVISIBLES (Marie Curie Actions, PITN- GA-2011- 289442).
\section*{}
\bibliographystyle{JHEP}
\bibliography{referencias_leptogenesis2}
\end{document}